\RequirePackage{amsmath}
\RequirePackage{flexisym}
\RequirePackage[superscript,biblabel,nomove]{cite}
\RequirePackage{graphicx}
\documentclass[12pt,aasms4]{article}
\usepackage{amsmath}
\usepackage{amssymb}
\usepackage{setspace}
\def\gsim{\;\rlap{\lower 2.5pt \hbox{$\sim$}}\raise 1.5pt\hbox{$>$}\;}
\def\lsim{\;\rlap{\lower 2.5pt  \hbox{$\sim$}}\raise 1.5pt\hbox{$<$}\;}
\newcommand\beq{\begin{equation}}
\newcommand\eeq{\end{equation}}

\begin{document}

\Large
\centerline{\bf Breakup of a Long-Period Comet as the}
\centerline{\bf Origin of the Dinosaur Extinction}

\medskip
\medskip
\normalsize
%\author{}
\centerline{ Amir Siraj$^{1\star}$ \& Abraham Loeb$^1$}
\medskip
\medskip
\medskip
\centerline{\it $^1$Department of Astronomy, Harvard University}
\centerline{\it 60 Garden Street, Cambridge, MA 02138, USA}
\medskip
\medskip
\small
\centerline{$^{\star}$Correspondence to amir.siraj@cfa.harvard.edu}
\normalsize
\vskip 0.2in
\hrule
\vskip 0.2in

\begin{doublespace}
{\bf The origin of the Chicxulub impactor, which is attributed as the cause of the K/T mass extinction event, is an unsolved puzzle.\cite{1980Sci...208.1095A, 2010Sci...327.1214S, 2007Natur.449...48B, 2009M&PS...44.1917R, 2011ApJ...741...68M} The background impact rates of main-belt asteroids and long-period comets have been previously dismissed as being too low to explain the Chicxulub impact event.\cite{2007Natur.449...48B} Here, we show that a fraction of long-period comets are tidally disrupted after passing close to the Sun, each producing a collection of smaller fragments that cross the orbit of Earth. This population could increase the impact rate of long-period comets capable of producing Chicxulub impact events by an order of magnitude. This new rate would be consistent with the age of the Chicxulub impact crater, thereby providing a satisfactory explanation for the origin of the impactor. Our hypothesis explains the composition of the largest confirmed impact crater in Earth's history\cite{2017E&PSL.460..105M} as well as the largest one within the last million years\cite{2017NatCo...8..227M}. It predicts a larger proportion of impactors with carbonaceous chondritic compositions than would be expected from meteorite falls of main-belt asteroids.}

%INTRO
Strong evidence suggests that the Chicxulub impact led to the K/T mass extinction event, which was the largest in the past $\sim 250 \mathrm{\; Myr}$ and brought about the demise of the dinosaurs.\cite{1980Sci...208.1095A, 2010Sci...327.1214S} However, the nature of the Chicxulub impactor is poorly understood. The latest scenario suggested postulated that the breakup of the Baptisina asteroid family could have led to the formation of the Chicxulub impactor.\cite{2007Natur.449...48B} However, spectroscopic follow-up indicated that the Baptistina family has an S-type, rather than an Xc-type composition, making it an unlikely source of the Chicxulub impactor, which had a carbonaceous chondritic composition,\cite{1998Natur.396..237K, 2006E&PSL.241..780T, 2009M&PS...44.1917R} although not ruling out entirely the possibility due to the stochastic nature of asteroid collisions and the subsequent disruptive processes.\cite{2006mess.book..829E} Observations of the Baptisina family also suggested that the breakup age may be $\sim 80 \mathrm{\; Myr}$ \cite{2011ApJ...741...68M} rather than $\sim 160 \mathrm{\; Myr}$ \cite{2007Natur.449...48B}, further reducing the likelihood that the Baptisina breakup formed the Chicxulub impactor.

%\newpage
%IMPACT RATES

The Chicxulub impactor could have originated from the background populations of asteroids or of comets. Main-belt asteroids (MBAs) with diameters $D \gtrsim 10 \mathrm{\; km}$, capable of producing Chicxulub impact events, strike the Earth once per $\sim 350 \mathrm{\; Myr}$.\cite{2002Icar..156..399B, 2018Icar..312..181G} Based on meteorite fall statistics,\cite{2002aste.book..653B} one such object with a carbonaceous chondritic composition impacts the Earth over a characteristic timescale of $\sim 3.5 \mathrm{\; Gyr}$, too rare to account for the K/T event.\cite{2007Natur.449...48B} Long-period comets (LPCs) capable of producing Chicxulub-scale impacts strike Earth also too rarely, once per $\sim 3.8 - 11 \mathrm{\; Gyr}$,\cite{2007Natur.449...48B} based on the rate of Earth-crossing LPCs and the impact probability per perihelion passage,\cite{2005ApJ...635.1348F, 2007IAUS..236..441W} and adopting a cumulative power-law index within the range -2.0 to -2.7.\cite{2006IAUS..229...67T, 2006Icar..185..211F, 2006Icar..182..527T} The only cometary sample-return mission to date, \textit{Stardust}, found that Comet 81P/Wild 2 had carbonaceous chondritic composition, suggesting that such a composition could potentially be widespread in comets.\cite{2008Sci...321.1664N, 2008M&PS...43..261Z, 2011PNAS..10819171C, 2012E&PSL.341..186B} As a result, the rate of LPC impacts with carbonaceous chondritic composition could be similar to the overall LPC impact rate. Within a timescale of $\sim 100 \mathrm{\; Myr}$, stellar encounters could boost the impactor flux by an order of magnitude for a Myr timescale,\cite{2019AJ....157..181V} which are insufficient in magnitude to explain a Chicxulub impact event. We note that comets are typically more fragile and porous than asteroids. \cite{2006ApJ...652.1768B, 2009P&SS...57..243T}

%SIMULATION

To find the fraction of LPCs with orbital behavior that could affect the impact flux at Earth, we simulated gravitational interactions between LPCs and the Jupiter-Earth-Sun system using a semi-analytic approach. Initially, there are $N$ Jupiter-crossing LPCs (initial pericenter distance $q \lesssim 5.2 \mathrm{\; AU}$) with semi-major axis $a \sim 10^4 \mathrm{\; AU}$ and the distribution of pericenter distances scaling as $q^2$, the corresponding cross-sectional area.\cite{2017A&A...604A..24F, 2019AJ....157..181V} The initial inclination distribution is taken as uniform.\cite{2017A&A...604A..24F, 2019AJ....157..181V} We then follow the orbital perturbation prescription for a restricted three-body scattering.\cite{2006tbp..book.....V} At the initial closest approach to Jupiter, calculated by selecting a random phase angle in Jupiter's orbit and computing the minimum distance between Jupiter and the LPC's orbit $b_{J}$, the change in semi-major axis $a$ resulting from the three-body interaction is computed as $\Delta (1/a) = (4 M_{J} v_{J} \sqrt{a} (\cos{\gamma} + K \cos{\delta}) / M_{\odot}^{3/2} b_J  \sqrt{G} (1 + K^2))$, where $M_J$ is the mass of Jupiter, $M_{\odot}$ is the mass of the Sun, $v_J$ is the heliocentric orbital speed of Jupiter, $G$ is the gravitational constant, $\gamma$ is the angle between the velocity vectors of Jupiter and the LPC, $\delta$ is the angle between the normal in the orbital plane to the approach of the LPC at the time of its closest approach to Jupiter and the velocity vector of Jupiter, and $K \equiv (G M_J a / M_{\odot} b_J)$. The new inclination is approximated by the numerically derived fitting function, $\approx \arccos{[\cos{i} - 0.38 \sin{i}^2 Q^{-1/2} (b_J / a)]}$, where $Q \equiv (q / b_J)$. The updated eccentricity is calculated through conservation of the Tisserand parameter, $T = (1/a) + 2 \sqrt{a(1 - e^2)} \cos{i}$, across the encounter. If the LPC crosses the orbit of Earth, defined as $q \lesssim 1 \mathrm{\; AU}$, the same process of updating the orbital is repeated for the closest encounter with the Earth, for a random Earth phase angle. We consider LPCs with $a > 2 \times 10^5 \mathrm{\; AU}$ or $e \geq 1$ to be ejected and remove them from the simulation as well as any that collide with Jupiter, the Sun, or the Earth. Tidal disruption by Jupiter is similar in likelihood to collision with Jupiter, $\sim 10^{-8}$ per Jupiter-crossing orbit.

We find that for $N = 10^5$ particles, $\sim 20\%$ of Earth-crossing events, defined as perihelia within the orbital radius of the Earth $q \lesssim 1 \mathrm{\; AU}$), were immediately preceded by perihelia within the Roche radius of the Sun, $q \lesssim r_{\odot} (2 \rho_{\odot} / \rho_{obj})^{1/3}$, where $r_{\odot}$ is the radius of the Sun, $\rho_{\odot}$ is the mean mass density of the Sun, and $\rho_{obj} \sim 0.7 \mathrm{\; g \; cm^{-3}}$ is the mean density\cite{2008M&PS...43.1033W} of the LPC, since they were captured into highly eccentric orbits by interacting with the Sun-Jupiter system. This is consistent with previous estimates of the sungrazing LPC population.\cite{1992A&A...257..315B} If the LPC is solely bound by gravity, then it is tidally disrupted. This is consistent with comets being the most fragile bodies in the Solar system, being mostly formed by weakly bound aggregates. \cite{2006ApJ...652.1768B, 2006Sci...314.1711B, 2006MNRAS.372..655T, 2015EGUGA..1710230T} Some comets may be highly heterogeneous rubble piles as a result of impact gardening and collisional processes,\cite{2009P&SS...57..243T, 2016ApJ...824...12B} with some pieces having relatively higher strengths, as was proposed to explain the origin of rare H/L chondrites.\cite{2006MNRAS.372..655T, 2017ASSP...46...11T} The characteristic change in $v_{\infty}$ for the fragments is, $\Delta v_{\infty} \sim \sqrt{v \Delta v}$, where $v \sim \sqrt{G M_{\odot}/d_{\odot, R}}$ and $\Delta v \sim \sqrt{G m/R}$, where $d_{\odot, R}$ is the Sun's Roche radius, $m$ is the mass of the progenitor, and $R$ is the radius of the progenitor. The change in $v_{\infty}$, $\Delta v_{\infty}$, is comparable to the original $v_{\infty}$ for an LPC. The time between disruption and crossing the Earth's orbit  is $\sim (d_{\oplus} / \sqrt{GM_{\odot} / d_{\oplus}}) \sim 10^3 \tau$, where $d_{\oplus} \sim 1 \mathrm{\; AU}$ is the distance of the Earth and $\tau$ is the tidal disruption encounter timescale, $\tau \equiv \sqrt{d_{\odot}^3 / G M_{\odot}}$. This is consistent with the conversion\cite{1998P&SS...46.1677H} of $R \sim 30 \mathrm{\; km}$ LPCs into fragments with effective radii of $R \sim 3.5 \mathrm{\; km}$, as required for the Chicxulub impactor, using a framework consistent with the Shoemaker-Levy 9 event\cite{2018ARA&A..56..593W} as well as the formation of the Gomul and Gipul crater chains. Data from Gomul and Gipul, as well other crater chains on Callisto and Ganymede, indicate that the fragments typically vary in size only by a factor of order unity\cite{1996Icar..121..249S}, due to the gravitationally bound rubble pile fragmentation model, although some second-order disruption effects are possible. We note that the canonical equation\cite{2005M&PS...40..817C} $z_b = z_{\star} - 2H \left[ \ln{ 1 + (l / 2H) \sqrt{f_p^2 - 1}} \right]$ for the parameters considered here is only consistent with $z_b < 0$, implying that despite experiencing disruption during atmospheric entry,\cite{2008IJIE...35.1441B} the comet fragment does not suffer an airburst, which was the fate of the Tunguska impactor,\cite{1978BAICz..29..129K, 1998P&SS...46..205A} but instead forms a crater, as observed. In the equation above, $z_b$ is the altitude at which the airburst occurs, $z_\star$ is the altitude at which the comet begins to disrupt, $H$ is the scale height of the atmosphere, $l = L_0 \sin(\theta) \sqrt{\rho_{obj} / (C_D \rho_a(z_\star)}$ is the dispersion length scale, $f_p =(L(z) / L_0)$ is the pancake factor, $L = 2R$ is the impactor diameter, $\rho_{obj}$ is the impactor density, $\rho_a$ is the atmospheric density, $\theta$ is the impact angle, and $C_D$ is a drag coefficient.

%ENHANCED IMPACT RATE

We now consider the effect that tidal disruption of a fraction of LPCs has on the impact rate of cometary bodies capable of producing Chicxulub. We first note that $D \gtrsim 10 \mathrm{\; km}$ progenitors, as considered here, are not thermally disrupted at large distances like smaller comets.\cite{2012MNRAS.423.1674F} We adopt the size distribution of Kuiper belt objects (KBOs) as a proxy for large LPCs or Oort cloud objects, due to their shared histories.\cite{2004come.book..175M, 2003Natur.424..639S, 2008ssbn.book..293K} KBOs with radii ranging from $R \sim 5 - 10 \mathrm{\; km}$ and $R \sim 30 \mathrm{\; km}$ can be described with a power-law index of $q \sim 2$ for a cumulative size distribution of the form,\cite{2009AJ....137...72F, 2013AJ....146...36S} $N(>R) \propto R^{1-q}$. The size distribution for LPCs, which have been observed up to radii of $R \sim 10 \mathrm{\; km}$, is consistent with the extrapolation of the $q \sim 2$ power law down to a the size of a cometary Chicxulub impactor,\cite{2012MNRAS.423.1674F, 2019Icar..333..252B} $R \sim 3.5 \mathrm{\; km}$. KBOs with $R \sim 30 \mathrm{\; km}$ are primarily bound by gravity, as indicated by modeling consistent with the observed size-density relationship\cite{2013ApJ...778L..34B, 2014A&A...570A..47W} and as implied by the location of the break in the size distribution.\cite{2004AJ....128.1916K, 2009AJ....137...72F, 2013AJ....146...36S} Most asteroids with sizes of $D \gtrsim 10 \mathrm{\; km}$ are not considered strengthless, meaning that if they passed within the Sun's Roche limit, they most likely would not produce fragments of the necessary size to explain Chicxulub.\cite{2018ARA&A..56..593W}

Since the mass of an LPC scales as $R^3$ and the abundance of LPCs scales as $R^{1-q}$, the overall enhancement of the time-averaged flux of cometary impactors capable of producing Chicxulub impact events resulting from the breakup and immediate crossing of the $\sim 1 \mathrm{\; AU}$ sphere following perihelion of objects larger than an intact LPC capable of producing a Chicxulub impact event by a factor of $\sim 10$ in radius is, $\sim 0.2 \times (30 \mathrm{\; km} / 3.5 \mathrm{\; km})^{3+(1-q)} \approx 15$, since 20\% of progenitors are tidally disrupted. This results in an impact rate for LPC fragments capable of producing Chicxulub impact events of once per $\sim 250 - 730 \mathrm{\; Myr}$. Irrespective of composition, the total impact rate of LPC fragments that could cause Chicxulub impact events is comparable to the total impact rate of MBAs that trigger events. We note that in order to be in agreement with the lack of an observed increase in the Earth's dust accretion rate across the K/T event over timescales of $\sim 1 \mathrm{\; Myr}$, the power-law index of the differential size distribution at the time of the tidally disrupted LPC's encounter with Earth must have been $q \gtrsim -3$, which can be tested through detailed modeling of such tidal disruption events.

%DISCUSSION

%Using KBOs as a proxy for the size distribution of large LPCs, we found that tidally disrupted KOCs with Chicxulub energies is once per $\sim 250 - 730 \mathrm{\; Myr}$.

The carbonaceous chondritic composition fraction of LPCs might be comparable to unity, since the first cometary target of a sample return mission Comet 81P/Wild 2 indicated a carbonaceous chondritic composition. However, the tiny aggregate particles collected had very low tensile strengths, potentially complicating the understanding of cometary structure in general.\cite{2006Sci...314.1711B} Adopting the assumption that the carbonaceous chondritic composition fraction of LPCs might be comparable to unity, the impact rate of tidally-disrupted LPCs is consistent with the Chicxulub impact event being the largest mass extinction event in the last $\sim 250 \mathrm{\; Myr}$, and is significantly larger than the impact rate of MBAs that could cause Chicxulub impact events. In particular, the probability that the Chicxulub impactor was an LPC fragment is larger than the probability that it was an MBA if the carbonaceous chondritic composition fraction of the LPC progenitors is $\gtrsim 7 - 20 \%$.

\begin{figure*}[hptb]
%\epsscale{.5}
\includegraphics{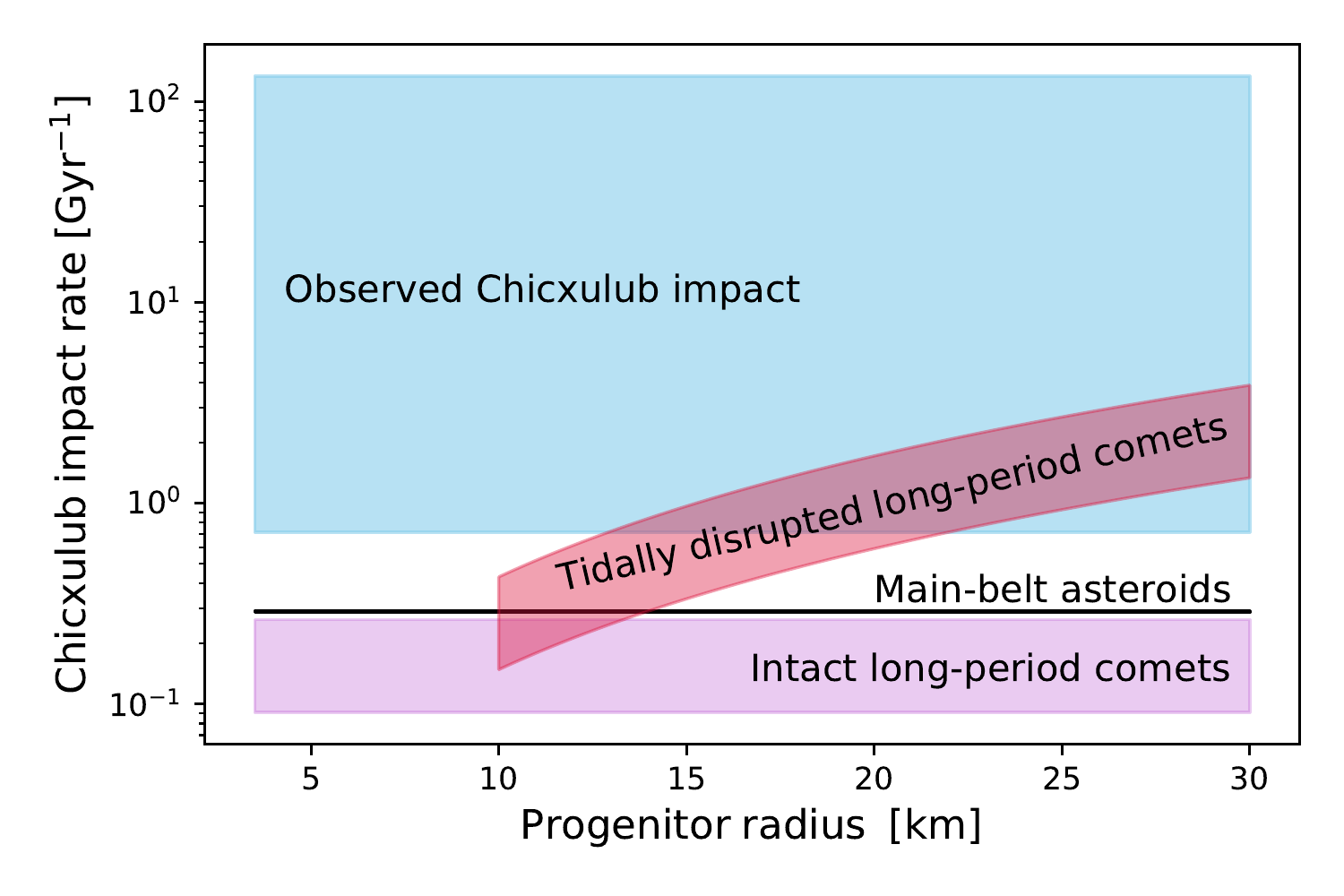}

\caption{\doublespacing The impact rate of tidally disrupted LPCs with energies comparable to that of the Chicxulub impactor, with the impact rates of intact LPCs and MBAs for reference, in addition to the range of rates that would explain the observed Chicxulub impact, including 95\% Poisson errors. Most LPCs and $\sim 10\%$ of MBAs are assumed to have a carbonaceous chondritic composition (see text for details).}
\end{figure*}

As illustrated in Figure 1, the LPC fragment hypothesis is consistent with the $95\%$ Poisson limits on the observed Chicxulub impact rate for progenitor carbonaceous chondritic composition fractions of $\gtrsim 20 - 50 \%$. Future cometary sample-return missions similar to \textit{Stardust} will constrain the fraction of comets with carbonaceous chondritic compositions and thereby serve as important test for our hypothesis. In addition, measurements of the size distribution of Oort cloud objects will improve the precision of our model. Since comets with $D \lesssim 10 \mathrm{\; km}$ are thermally disrupted at large distances from the Sun\cite{2012MNRAS.423.1674F} and also the size distribution of comets with $D \gtrsim 60 \mathrm{\; km}$ is described by a power law\cite{2013AJ....146...36S} with a cumulative power-law index steeper than -3, our model only applies to the progenitor size range of $10 \mathrm{\; km} \lesssim D \lesssim 60 \mathrm{\; km}$, thereby not affecting the overall crater size distribution.

Our hypothesis predicts that other Chicxulub-size craters on Earth are more likely to correspond to an impactor with a carbonaceous chondritc composition than expected from the carbonaceous chondritc composition fraction of MBAs. We note that meteorite fall statistics should still reflect the compositions of asteroids, as canonically assumed. For small LPCs that pass within the Sun's Roche radius, the ablated mass is $\sim (R^2 L_{\odot} \tau / 8 d_{\odot}^2 Q)$, where $L_{\odot}$ is the luminosity of the Sun, $d_{\odot, R}$ is the Roche radius of the Sun, $\tau$ is the encounter timescale, and $Q$ is the energy per unit mass necessary to vaporize the material. Adopting\cite{2012MNRAS.421.1315Z} $Q \sim 3 \times 10^{11} \mathrm{\; erg \; g^{-1}}$, the initial mass is comparable to the ablated mass for object radii of $R \sim 1 \mathrm{\; m}$, resulting in a conservative lower bound on the mass of LPC fragments of $\sim 10^5 \mathrm{\; g}$, which is orders of magnitude above the preatmospheric entry masses of objects that dominate the meteorite flux at the Earth's surface.\cite{2006mess.book..869Z} This magnitude of ablation indicates that mass loss is negligible for the progenitor size range considered here. In addition, the heating due to solar irradiation, $\sim 10^3 \mathrm{\; K}$ over  $\sim 10^3 \mathrm{\; s}$, does not exceed the expected heating from the impact itself,\cite{2005M&PS...40..817C} so no additional signatures of thermal processing would be expected. Shoemaker-Levy 9, 2015 TB145, and the Encke complex are all examples of large fragments resulting from tidal disruption.\cite{2008MNRAS.386.1436B, 2015EGUGA..1710230T, 2017A&A...598A..63M} Additionally, the observation that the largest particles in most observed meteoroid streams are cm-sized\cite{1998EP&S...50..555J} is not surprising, since larger particles are naturally more rare than smaller particles.

Indeed, Vredefort, the only confirmed crater on Earth larger than Chicxulub (by a factor of $\sim 2$ in radius),\cite{2005SoSyR..39..381I}  may correspond to an impactor with a carbonaceous chondritic composition.\cite{2017E&PSL.460..105M} Additionally, since LPC fragment Chicxulub impactors should strike Earth once every $\sim 250 - 730 \mathrm{\; Myr}$, fragments an order of magnitude smaller in radius, if produced by the same progenitors, would strike Earth no more frequently than once per $\sim 0.25 - 0.73 \mathrm{\; Myr}$ and if a significant fraction of the progenitors have a carbonaceous chondritc composition, the most recent such crater should reflect such a composition. Indeed, the Zhamanshin crater, the largest confirmed impact crater on Earth formed in the last $\sim \mathrm{Myr}$ (an order of magnitude smaller in radius than Chicxulub),\cite{2020GeCoA.268..209S} shows evidence that the impactor may have had a carbonaceous chondritc composition,\cite{2017NatCo...8..227M} providing support to our model. Additionally, the likely existence of a well-separated reservoir of carbonaceous chondritic material beyond the orbit of Jupiter in the solar protoplanetary disk\cite{2020arXiv200613528S} lends further support to our model. Our model is in no conflict with the Moon's cratering rate, since it only applies in the size range around Chicxulub-scale impactors. The cross-sectional area of the Moon is an order of magnitude smaller than Earth, implying that a Chicxulub size impactor would be very rare (once per few Gyr), and thereby implying that such an LPC impact event may have not happened for the Moon.

\end{doublespace}

\vskip 0.45in
\hrule
\vskip 0.15in

\small
\noindent

%\bibliographystyle{naturemag}
%\bibliography{bib} %{}

%\bibliographystyle{nature}
%\citestyle{nature}
%\bibliography{bib}

%\printbibliography
%\bibliography{nature/p.bib}

\begin{thebibliography}{10}
\expandafter\ifx\csname url\endcsname\relax
  \def\url#1{\texttt{#1}}\fi
\expandafter\ifx\csname urlprefix\endcsname\relax\def\urlprefix{URL }\fi
\providecommand{\bibinfo}[2]{#2}
\providecommand{\eprint}[2][]{\url{#2}}

\bibitem{1980Sci...208.1095A}
\bibinfo{author}{{Alvarez}, L.~W.}, \bibinfo{author}{{Alvarez}, W.},
  \bibinfo{author}{{Asaro}, F.} \& \bibinfo{author}{{Michel}, H.~V.}
\newblock \bibinfo{title}{{Extraterrestrial Cause for the Cretaceous-Tertiary
  Extinction}}.
\newblock \emph{\bibinfo{journal}{Science}} \textbf{\bibinfo{volume}{208}},
  \bibinfo{pages}{1095--1108} (\bibinfo{year}{1980}).

\bibitem{2010Sci...327.1214S}
\bibinfo{author}{{Schulte}, P.} \emph{et~al.}
\newblock \bibinfo{title}{{The Chicxulub Asteroid Impact and Mass Extinction at
  the Cretaceous-Paleogene Boundary}}.
\newblock \emph{\bibinfo{journal}{Science}} \textbf{\bibinfo{volume}{327}},
  \bibinfo{pages}{1214} (\bibinfo{year}{2010}).

\bibitem{2007Natur.449...48B}
\bibinfo{author}{{Bottke}, W.~F.}, \bibinfo{author}{{Vokrouhlick{\'y}}, D.} \&
  \bibinfo{author}{{Nesvorn{\'y}}, D.}
\newblock \bibinfo{title}{{An asteroid breakup 160Myr ago as the probable
  source of the K/T impactor}}.
\newblock \emph{\bibinfo{journal}{Nature}} \textbf{\bibinfo{volume}{449}},
  \bibinfo{pages}{48--53} (\bibinfo{year}{2007}).

\bibitem{2009M&PS...44.1917R}
\bibinfo{author}{{Reddy}, V.} \emph{et~al.}
\newblock \bibinfo{title}{{Composition of 298 Baptistina: Implications for the
  K/T impactor link}}.
\newblock \emph{\bibinfo{journal}{Meteoritics and Planetary Science}}
  \textbf{\bibinfo{volume}{44}}, \bibinfo{pages}{1917--1927}
  (\bibinfo{year}{2009}).

\bibitem{2011ApJ...741...68M}
\bibinfo{author}{{Masiero}, J.~R.} \emph{et~al.}
\newblock \bibinfo{title}{{Main Belt Asteroids with WISE/NEOWISE. I.
  Preliminary Albedos and Diameters}}.
\newblock \emph{\bibinfo{journal}{ApJ}} \textbf{\bibinfo{volume}{741}},
  \bibinfo{pages}{68} (\bibinfo{year}{2011}).
\newblock \eprint{1109.4096}.

\bibitem{2017E&PSL.460..105M}
\bibinfo{author}{{Mougel}, B.}, \bibinfo{author}{{Moynier}, F.},
  \bibinfo{author}{{G{\"o}pel}, C.} \& \bibinfo{author}{{Koeberl}, C.}
\newblock \bibinfo{title}{{Chromium isotope evidence in ejecta deposits for the
  nature of Paleoproterozoic impactors}}.
\newblock \emph{\bibinfo{journal}{Earth and Planetary Science Letters}}
  \textbf{\bibinfo{volume}{460}}, \bibinfo{pages}{105--111}
  (\bibinfo{year}{2017}).
\newblock \eprint{1612.06922}.

\bibitem{2017NatCo...8..227M}
\bibinfo{author}{{Magna}, T.} \emph{et~al.}
\newblock \bibinfo{title}{{Zhamanshin astrobleme provides evidence for
  carbonaceous chondrite and post-impact exchange between ejecta and Earth's
  atmosphere}}.
\newblock \emph{\bibinfo{journal}{Nature Communications}}
  \textbf{\bibinfo{volume}{8}}, \bibinfo{pages}{227} (\bibinfo{year}{2017}).

\bibitem{1998Natur.396..237K}
\bibinfo{author}{{Kyte}, F.~T.}
\newblock \bibinfo{title}{{A meteorite from the Cretaceous/Tertiary boundary}}.
\newblock \emph{\bibinfo{journal}{Nature}} \textbf{\bibinfo{volume}{396}},
  \bibinfo{pages}{237--239} (\bibinfo{year}{1998}).

\bibitem{2006E&PSL.241..780T}
\bibinfo{author}{{Trinquier}, A.}, \bibinfo{author}{{Birck}, J.-L.} \&
  \bibinfo{author}{{Jean All{\`e}gre}, C.}
\newblock \bibinfo{title}{{The nature of the KT impactor. A $^{54}$Cr
  reappraisal}}.
\newblock \emph{\bibinfo{journal}{Earth and Planetary Science Letters}}
  \textbf{\bibinfo{volume}{241}}, \bibinfo{pages}{780--788}
  (\bibinfo{year}{2006}).

\bibitem{2006mess.book..829E}
\bibinfo{author}{{Eugster}, O.}, \bibinfo{author}{{Herzog}, G.~F.},
  \bibinfo{author}{{Marti}, K.} \& \bibinfo{author}{{Caffee}, M.~W.}
\newblock \emph{\bibinfo{title}{{Irradiation Records, Cosmic-Ray Exposure Ages,
  and Transfer Times of Meteorites}}}, \bibinfo{pages}{829}
  (\bibinfo{year}{2006}).

\bibitem{2002Icar..156..399B}
\bibinfo{author}{{Bottke}, W.~F.} \emph{et~al.}
\newblock \bibinfo{title}{{Debiased Orbital and Absolute Magnitude Distribution
  of the Near-Earth Objects}}.
\newblock \emph{\bibinfo{journal}{Icarus}} \textbf{\bibinfo{volume}{156}},
  \bibinfo{pages}{399--433} (\bibinfo{year}{2002}).

\bibitem{2018Icar..312..181G}
\bibinfo{author}{{Granvik}, M.} \emph{et~al.}
\newblock \bibinfo{title}{{Debiased orbit and absolute-magnitude distributions
  for near-Earth objects}}.
\newblock \emph{\bibinfo{journal}{Icarus}} \textbf{\bibinfo{volume}{312}},
  \bibinfo{pages}{181--207} (\bibinfo{year}{2018}).
\newblock \eprint{1804.10265}.

\bibitem{2002aste.book..653B}
\bibinfo{author}{{Burbine}, T.~H.}, \bibinfo{author}{{McCoy}, T.~J.},
  \bibinfo{author}{{Meibom}, A.}, \bibinfo{author}{{Gladman}, B.} \&
  \bibinfo{author}{{Keil}, K.}
\newblock \emph{\bibinfo{title}{{Meteoritic Parent Bodies: Their Number and
  Identification}}}, \bibinfo{pages}{653--667} (\bibinfo{year}{2002}).

\bibitem{2005ApJ...635.1348F}
\bibinfo{author}{{Francis}, P.~J.}
\newblock \bibinfo{title}{{The Demographics of Long-Period Comets}}.
\newblock \emph{\bibinfo{journal}{ApJ}} \textbf{\bibinfo{volume}{635}},
  \bibinfo{pages}{1348--1361} (\bibinfo{year}{2005}).
\newblock \eprint{astro-ph/0509074}.

\bibitem{2007IAUS..236..441W}
\bibinfo{author}{{Weissman}, P.~R.}
\newblock \bibinfo{title}{{The cometary impactor flux at the Earth}}.
\newblock In \bibinfo{editor}{{Valsecchi}, G.~B.},
  \bibinfo{editor}{{Vokrouhlick{\'y}}, D.} \& \bibinfo{editor}{{Milani}, A.}
  (eds.) \emph{\bibinfo{booktitle}{Near Earth Objects, our Celestial Neighbors:
  Opportunity and Risk}}, vol. \bibinfo{volume}{236} of
  \emph{\bibinfo{series}{IAU Symposium}}, \bibinfo{pages}{441--450}
  (\bibinfo{year}{2007}).

\bibitem{2006IAUS..229...67T}
\bibinfo{author}{{Toth}, I.}
\newblock \bibinfo{title}{{Connections between asteroids and cometary nuclei}}.
\newblock In \bibinfo{editor}{{Lazzaro}, D.}, \bibinfo{editor}{{Ferraz-Mello},
  S.} \& \bibinfo{editor}{{Fern{\'a}ndez}, J.~A.} (eds.)
  \emph{\bibinfo{booktitle}{Asteroids, Comets, Meteors}}, vol.
  \bibinfo{volume}{229} of \emph{\bibinfo{series}{IAU Symposium}},
  \bibinfo{pages}{67--96} (\bibinfo{year}{2006}).

\bibitem{2006Icar..185..211F}
\bibinfo{author}{{Fern{\'a}ndez}, J.~A.} \& \bibinfo{author}{{Morbidelli}, A.}
\newblock \bibinfo{title}{{The population of faint Jupiter family comets near
  the Earth}}.
\newblock \emph{\bibinfo{journal}{Icarus}} \textbf{\bibinfo{volume}{185}},
  \bibinfo{pages}{211--222} (\bibinfo{year}{2006}).

\bibitem{2006Icar..182..527T}
\bibinfo{author}{{Tancredi}, G.}, \bibinfo{author}{{Fern{\'a}ndez}, J.~A.},
  \bibinfo{author}{{Rickman}, H.} \& \bibinfo{author}{{Licandro}, J.}
\newblock \bibinfo{title}{{Nuclear magnitudes and the size distribution of
  Jupiter family comets}}.
\newblock \emph{\bibinfo{journal}{Icarus}} \textbf{\bibinfo{volume}{182}},
  \bibinfo{pages}{527--549} (\bibinfo{year}{2006}).

\bibitem{2008Sci...321.1664N}
\bibinfo{author}{{Nakamura}, T.} \emph{et~al.}
\newblock \bibinfo{title}{{Chondrulelike Objects in Short-Period Comet 81P/Wild
  2}}.
\newblock \emph{\bibinfo{journal}{Science}} \textbf{\bibinfo{volume}{321}},
  \bibinfo{pages}{1664} (\bibinfo{year}{2008}).

\bibitem{2008M&PS...43..261Z}
\bibinfo{author}{{Zolensky}, M.} \emph{et~al.}
\newblock \bibinfo{title}{{Comparing Wild 2 particles to chondrites and IDPs}}.
\newblock \emph{\bibinfo{journal}{Meteoritics and Planetary Science}}
  \textbf{\bibinfo{volume}{43}}, \bibinfo{pages}{261--272}
  (\bibinfo{year}{2008}).

\bibitem{2011PNAS..10819171C}
\bibinfo{author}{{Cody}, G.~D.} \emph{et~al.}
\newblock \bibinfo{title}{{Cosmochemistry Special Feature: Establishing a
  molecular relationship between chondritic and cometary organic solids}}.
\newblock \emph{\bibinfo{journal}{Proceedings of the National Academy of
  Science}} \textbf{\bibinfo{volume}{108}}, \bibinfo{pages}{19171--19176}
  (\bibinfo{year}{2011}).

\bibitem{2012E&PSL.341..186B}
\bibinfo{author}{{Bridges}, J.~C.}, \bibinfo{author}{{Changela}, H.~G.},
  \bibinfo{author}{{Nayakshin}, S.}, \bibinfo{author}{{Starkey}, N.~A.} \&
  \bibinfo{author}{{Franchi}, I.~A.}
\newblock \bibinfo{title}{{Chondrule fragments from Comet Wild2: Evidence for
  high temperature processing in the outer Solar System}}.
\newblock \emph{\bibinfo{journal}{Earth and Planetary Science Letters}}
  \textbf{\bibinfo{volume}{341}}, \bibinfo{pages}{186--194}
  (\bibinfo{year}{2012}).

\bibitem{2019AJ....157..181V}
\bibinfo{author}{{Vokrouhlick{\'y}}, D.}, \bibinfo{author}{{Nesvorn{\'y}}, D.}
  \& \bibinfo{author}{{Dones}, L.}
\newblock \bibinfo{title}{{Origin and Evolution of Long-period Comets}}.
\newblock \emph{\bibinfo{journal}{AJ}} \textbf{\bibinfo{volume}{157}},
  \bibinfo{pages}{181} (\bibinfo{year}{2019}).
\newblock \eprint{1904.00728}.

\bibitem{2006ApJ...652.1768B}
\bibinfo{author}{{Blum}, J.}, \bibinfo{author}{{Schr{\"a}pler}, R.},
  \bibinfo{author}{{Davidsson}, B. J.~R.} \&
  \bibinfo{author}{{Trigo-Rodr{\'\i}guez}, J.~M.}
\newblock \bibinfo{title}{{The Physics of Protoplanetesimal Dust Agglomerates.
  I. Mechanical Properties and Relations to Primitive Bodies in the Solar
  System}}.
\newblock \emph{\bibinfo{journal}{ApJ}} \textbf{\bibinfo{volume}{652}},
  \bibinfo{pages}{1768--1781} (\bibinfo{year}{2006}).

\bibitem{2009P&SS...57..243T}
\bibinfo{author}{{Trigo-Rodriguez}, J.~M.} \& \bibinfo{author}{{Blum}, J.}
\newblock \bibinfo{title}{{Tensile strength as an indicator of the degree of
  primitiveness of undifferentiated bodies}}.
\newblock \emph{\bibinfo{journal}{P\&SS}} \textbf{\bibinfo{volume}{57}},
  \bibinfo{pages}{243--249} (\bibinfo{year}{2009}).

\bibitem{2017A&A...604A..24F}
\bibinfo{author}{{Fouchard}, M.}, \bibinfo{author}{{Rickman}, H.},
  \bibinfo{author}{{Froeschl{\'e}}, C.} \& \bibinfo{author}{{Valsecchi}, G.~B.}
\newblock \bibinfo{title}{{Distribution of long-period comets: comparison
  between simulations and observations}}.
\newblock \emph{\bibinfo{journal}{Astronomy \& Astrophysics}}
  \textbf{\bibinfo{volume}{604}}, \bibinfo{pages}{A24} (\bibinfo{year}{2017}).

\bibitem{2006tbp..book.....V}
\bibinfo{author}{{Valtonen}, M.} \& \bibinfo{author}{{Karttunen}, H.}
\newblock \emph{\bibinfo{title}{{The Three-Body Problem}}}
  (\bibinfo{year}{2006}).

\bibitem{2008M&PS...43.1033W}
\bibinfo{author}{{Weissman}, P.~R.} \& \bibinfo{author}{{Lowry}, S.~C.}
\newblock \bibinfo{title}{{Structure and density of cometary nuclei}}.
\newblock \emph{\bibinfo{journal}{Meteoritics and Planetary Science}}
  \textbf{\bibinfo{volume}{43}}, \bibinfo{pages}{1033--1047}
  (\bibinfo{year}{2008}).

\bibitem{1992A&A...257..315B}
\bibinfo{author}{{Bailey}, M.~E.}, \bibinfo{author}{{Chambers}, J.~E.} \&
  \bibinfo{author}{{Hahn}, G.}
\newblock \bibinfo{title}{{Origin of sungrazers - A frequent cometary
  end-state.}}
\newblock \emph{\bibinfo{journal}{Astronomy \& Astrophysics}}
  \textbf{\bibinfo{volume}{257}}, \bibinfo{pages}{315--322}
  (\bibinfo{year}{1992}).

\bibitem{2006Sci...314.1711B}
\bibinfo{author}{{Brownlee}, D.} \emph{et~al.}
\newblock \bibinfo{title}{{Comet 81P/Wild 2 Under a Microscope}}.
\newblock \emph{\bibinfo{journal}{Science}} \textbf{\bibinfo{volume}{314}},
  \bibinfo{pages}{1711} (\bibinfo{year}{2006}).

\bibitem{2006MNRAS.372..655T}
\bibinfo{author}{{Trigo-Rodr{\'\i}guez}, J.~M.} \& \bibinfo{author}{{Llorca},
  J.}
\newblock \bibinfo{title}{{The strength of cometary meteoroids: clues to the
  structure and evolution of comets}}.
\newblock \emph{\bibinfo{journal}{MNRAS}} \textbf{\bibinfo{volume}{372}},
  \bibinfo{pages}{655--660} (\bibinfo{year}{2006}).

\bibitem{2015EGUGA..1710230T}
\bibinfo{author}{{Trigo-Rodr{\'\i}guez}, J.~M.}, \bibinfo{author}{{Rimola}, A.}
  \& \bibinfo{author}{{Martins}, Z.}
\newblock \bibinfo{title}{{Aqueous processing of organic compounds in
  carbonaceous asteroids}}.
\newblock In \emph{\bibinfo{booktitle}{EGU General Assembly Conference
  Abstracts}}, EGU General Assembly Conference Abstracts,
  \bibinfo{pages}{10230} (\bibinfo{year}{2015}).

\bibitem{2016ApJ...824...12B}
\bibinfo{author}{{Beitz}, E.}, \bibinfo{author}{{Blum}, J.},
  \bibinfo{author}{{Parisi}, M.~G.} \& \bibinfo{author}{{Trigo-Rodriguez}, J.}
\newblock \bibinfo{title}{{The Collisional Evolution of Undifferentiated
  Asteroids and the Formation of Chondritic Meteoroids}}.
\newblock \emph{\bibinfo{journal}{ApJ}} \textbf{\bibinfo{volume}{824}},
  \bibinfo{pages}{12} (\bibinfo{year}{2016}).
\newblock \eprint{1604.02340}.

\bibitem{2017ASSP...46...11T}
\bibinfo{author}{{Trigo-Rodr{\'\i}guez}, J.~M.} \& \bibinfo{author}{{Williams},
  I.~P.}
\newblock \bibinfo{title}{{Dynamic Sources of Contemporary Hazard from
  Meteoroids and Small Asteroids}}.
\newblock In \bibinfo{editor}{{Trigo-Rodr{\'\i}guez}, J.~M.},
  \bibinfo{editor}{{Gritsevich}, M.} \& \bibinfo{editor}{{Palme}, H.} (eds.)
  \emph{\bibinfo{booktitle}{Assessment and Mitigation of Asteroid Impact
  Hazards: Proceedings of the 2015 Barcelona Asteroid Day}},
  vol.~\bibinfo{volume}{46}, \bibinfo{pages}{11} (\bibinfo{year}{2017}).

\bibitem{1998P&SS...46.1677H}
\bibinfo{author}{{Hahn}, J.~M.} \& \bibinfo{author}{{Rettig}, T.~W.}
\newblock \bibinfo{title}{{Tidal disruption of strengthless rubble piles - a
  dimensional analysis}}.
\newblock \emph{\bibinfo{journal}{Planet. Space Sci.}}
  \textbf{\bibinfo{volume}{46}}, \bibinfo{pages}{1677--1682}
  (\bibinfo{year}{1998}).

\bibitem{2018ARA&A..56..593W}
\bibinfo{author}{{Walsh}, K.~J.}
\newblock \bibinfo{title}{{Rubble Pile Asteroids}}.
\newblock \emph{\bibinfo{journal}{Annual Review of Astronomy and Astrophysics}}
  \textbf{\bibinfo{volume}{56}}, \bibinfo{pages}{593--624}
  (\bibinfo{year}{2018}).
\newblock \eprint{1810.01815}.

\bibitem{1996Icar..121..249S}
\bibinfo{author}{{Schenk}, P.~M.}, \bibinfo{author}{{Asphaug}, E.},
  \bibinfo{author}{{McKinnon}, W.~B.}, \bibinfo{author}{{Melosh}, H.~J.} \&
  \bibinfo{author}{{Weissman}, P.~R.}
\newblock \bibinfo{title}{{Cometary Nuclei and Tidal Disruption: The Geologic
  Record of Crater Chains on Callisto and Ganymede}}.
\newblock \emph{\bibinfo{journal}{Icarus}} \textbf{\bibinfo{volume}{121}},
  \bibinfo{pages}{249--274} (\bibinfo{year}{1996}).

\bibitem{2005M&PS...40..817C}
\bibinfo{author}{{Collins}, G.~S.}, \bibinfo{author}{{Melosh}, H.~J.} \&
  \bibinfo{author}{{Marcus}, R.~A.}
\newblock \bibinfo{title}{{Earth Impact Effects Program: A Web-based computer
  program for calculating the regional environmental consequences of a
  meteoroid impact on Earth}}.
\newblock \emph{\bibinfo{journal}{Meteoritics and Planetary Science}}
  \textbf{\bibinfo{volume}{40}}, \bibinfo{pages}{817} (\bibinfo{year}{2005}).

\bibitem{2008IJIE...35.1441B}
\bibinfo{author}{{Boslough}, M.~B.~E.} \& \bibinfo{author}{{Crawford}, D.~A.}
\newblock \bibinfo{title}{{Low-Altitude Airbursts and the Impact Threat}}.
\newblock \emph{\bibinfo{journal}{International Journal of Impact Engineering}}
  \textbf{\bibinfo{volume}{35}}, \bibinfo{pages}{1441--1448}
  (\bibinfo{year}{2008}).

\bibitem{1978BAICz..29..129K}
\bibinfo{author}{{Kresak}, L.}
\newblock \bibinfo{title}{{The Tunguska Object: a Fragment of Comet Encke?}}
\newblock \emph{\bibinfo{journal}{Bulletin of the Astronomical Institutes of
  Czechoslovakia}} \textbf{\bibinfo{volume}{29}}, \bibinfo{pages}{129}
  (\bibinfo{year}{1978}).

\bibitem{1998P&SS...46..205A}
\bibinfo{author}{{Asher}, D.~J.} \& \bibinfo{author}{{Steel}, D.~I.}
\newblock \bibinfo{title}{{On the possible relation between the Tunguska bolide
  and comet Encke}}.
\newblock \emph{\bibinfo{journal}{P\&SS}} \textbf{\bibinfo{volume}{46}},
  \bibinfo{pages}{205--211} (\bibinfo{year}{1998}).

\bibitem{2012MNRAS.423.1674F}
\bibinfo{author}{{Fern{\'a}ndez}, J.~A.} \& \bibinfo{author}{{Sosa}, A.}
\newblock \bibinfo{title}{{Magnitude and size distribution of long-period
  comets in Earth-crossing or approaching orbits}}.
\newblock \emph{\bibinfo{journal}{MNRAS}} \textbf{\bibinfo{volume}{423}},
  \bibinfo{pages}{1674--1690} (\bibinfo{year}{2012}).
\newblock \eprint{1204.2285}.

\bibitem{2004come.book..175M}
\bibinfo{author}{{Morbidelli}, A.} \& \bibinfo{author}{{Brown}, M.~E.}
\newblock \emph{\bibinfo{title}{{The kuiper belt and the primordial evolution
  of the solar system}}}, \bibinfo{pages}{175} (\bibinfo{year}{2004}).

\bibitem{2003Natur.424..639S}
\bibinfo{author}{{Stern}, S.~A.}
\newblock \bibinfo{title}{{The evolution of comets in the Oort cloud and Kuiper
  belt}}.
\newblock \emph{\bibinfo{journal}{Nature}} \textbf{\bibinfo{volume}{424}},
  \bibinfo{pages}{639--642} (\bibinfo{year}{2003}).

\bibitem{2008ssbn.book..293K}
\bibinfo{author}{{Kenyon}, S.~J.}, \bibinfo{author}{{Bromley}, B.~C.},
  \bibinfo{author}{{O'Brien}, D.~P.} \& \bibinfo{author}{{Davis}, D.~R.}
\newblock \emph{\bibinfo{title}{{Formation and Collisional Evolution of Kuiper
  Belt Objects}}}, \bibinfo{pages}{293} (\bibinfo{year}{2008}).

\bibitem{2009AJ....137...72F}
\bibinfo{author}{{Fraser}, W.~C.} \& \bibinfo{author}{{Kavelaars}, J.~J.}
\newblock \bibinfo{title}{{The Size Distribution of Kuiper Belt Objects for D
  gsim 10 km}}.
\newblock \emph{\bibinfo{journal}{AJ}} \textbf{\bibinfo{volume}{137}},
  \bibinfo{pages}{72--82} (\bibinfo{year}{2009}).
\newblock \eprint{0810.2296}.

\bibitem{2013AJ....146...36S}
\bibinfo{author}{{Schlichting}, H.~E.}, \bibinfo{author}{{Fuentes}, C.~I.} \&
  \bibinfo{author}{{Trilling}, D.~E.}
\newblock \bibinfo{title}{{Initial Planetesimal Sizes and the Size Distribution
  of Small Kuiper Belt Objects}}.
\newblock \emph{\bibinfo{journal}{AJ}} \textbf{\bibinfo{volume}{146}},
  \bibinfo{pages}{36} (\bibinfo{year}{2013}).
\newblock \eprint{1301.7433}.

\bibitem{2019Icar..333..252B}
\bibinfo{author}{{Boe}, B.} \emph{et~al.}
\newblock \bibinfo{title}{{The orbit and size-frequency distribution of long
  period comets observed by Pan-STARRS1}}.
\newblock \emph{\bibinfo{journal}{Icarus}} \textbf{\bibinfo{volume}{333}},
  \bibinfo{pages}{252--272} (\bibinfo{year}{2019}).
\newblock \eprint{1905.13458}.

\bibitem{2013ApJ...778L..34B}
\bibinfo{author}{{Brown}, M.~E.}
\newblock \bibinfo{title}{{The Density of Mid-sized Kuiper Belt Object 2002
  UX25 and the Formation of the Dwarf Planets}}.
\newblock \emph{\bibinfo{journal}{ApJL}} \textbf{\bibinfo{volume}{778}},
  \bibinfo{pages}{L34} (\bibinfo{year}{2013}).
\newblock \eprint{1311.0553}.

\bibitem{2014A&A...570A..47W}
\bibinfo{author}{{Wahlberg Jansson}, K.} \& \bibinfo{author}{{Johansen}, A.}
\newblock \bibinfo{title}{{Formation of pebble-pile planetesimals}}.
\newblock \emph{\bibinfo{journal}{Astronomy \& Astrophysics}}
  \textbf{\bibinfo{volume}{570}}, \bibinfo{pages}{A47} (\bibinfo{year}{2014}).
\newblock \eprint{1408.2535}.

\bibitem{2004AJ....128.1916K}
\bibinfo{author}{{Kenyon}, S.~J.} \& \bibinfo{author}{{Bromley}, B.~C.}
\newblock \bibinfo{title}{{The Size Distribution of Kuiper Belt Objects}}.
\newblock \emph{\bibinfo{journal}{AJ}} \textbf{\bibinfo{volume}{128}},
  \bibinfo{pages}{1916--1926} (\bibinfo{year}{2004}).
\newblock \eprint{astro-ph/0406556}.

\bibitem{2012MNRAS.421.1315Z}
\bibinfo{author}{{Zubovas}, K.}, \bibinfo{author}{{Nayakshin}, S.} \&
  \bibinfo{author}{{Markoff}, S.}
\newblock \bibinfo{title}{{Sgr A* flares: tidal disruption of asteroids and
  planets?}}
\newblock \emph{\bibinfo{journal}{MNRAS}} \textbf{\bibinfo{volume}{421}},
  \bibinfo{pages}{1315--1324} (\bibinfo{year}{2012}).
\newblock \eprint{1110.6872}.

\bibitem{2006mess.book..869Z}
\bibinfo{author}{{Zolensky}, M.}, \bibinfo{author}{{Bland}, P.},
  \bibinfo{author}{{Brown}, P.} \& \bibinfo{author}{{Halliday}, I.}
\newblock \emph{\bibinfo{title}{{Flux of Extraterrestrial Materials}}},
  \bibinfo{pages}{869} (\bibinfo{year}{2006}).

\bibitem{2008MNRAS.386.1436B}
\bibinfo{author}{{Babadzhanov}, P.~B.}, \bibinfo{author}{{Williams}, I.~P.} \&
  \bibinfo{author}{{Kokhirova}, G.~I.}
\newblock \bibinfo{title}{{Near-Earth Objects in the Taurid complex}}.
\newblock \emph{\bibinfo{journal}{MNRAS}} \textbf{\bibinfo{volume}{386}},
  \bibinfo{pages}{1436--1442} (\bibinfo{year}{2008}).

\bibitem{2017A&A...598A..63M}
\bibinfo{author}{{M{\"u}ller}, T.~G.} \emph{et~al.}
\newblock \bibinfo{title}{{Large Halloween asteroid at lunar distance}}.
\newblock \emph{\bibinfo{journal}{A\&A}} \textbf{\bibinfo{volume}{598}},
  \bibinfo{pages}{A63} (\bibinfo{year}{2017}).
\newblock \eprint{1610.08267}.

\bibitem{1998EP&S...50..555J}
\bibinfo{author}{{Jenniskens}, P.}
\newblock \bibinfo{title}{{On the dynamics of meteoroid streams}}.
\newblock \emph{\bibinfo{journal}{Earth, Planets, and Space}}
  \textbf{\bibinfo{volume}{50}}, \bibinfo{pages}{555--567}
  (\bibinfo{year}{1998}).

\bibitem{2005SoSyR..39..381I}
\bibinfo{author}{{Ivanov}, B.~A.}
\newblock \bibinfo{title}{{Numerical Modeling of the Largest Terrestrial
  Meteorite Craters}}.
\newblock \emph{\bibinfo{journal}{Solar System Research}}
  \textbf{\bibinfo{volume}{39}}, \bibinfo{pages}{381--409}
  (\bibinfo{year}{2005}).

\bibitem{2020GeCoA.268..209S}
\bibinfo{author}{{Schulz}, T.} \emph{et~al.}
\newblock \bibinfo{title}{{The Zhamanshin impact structure, Kazakhstan: A
  comparative geochemical study of target rocks and impact glasses}}.
\newblock \emph{\bibinfo{journal}{GCA}} \textbf{\bibinfo{volume}{268}},
  \bibinfo{pages}{209--229} (\bibinfo{year}{2020}).

\bibitem{2020arXiv200613528S}
\bibinfo{author}{{Spitzer}, F.} \emph{et~al.}
\newblock \bibinfo{title}{{Isotopic evolution of the inner Solar System
  inferred from molybdenum isotopes in meteorites}}.
\newblock \emph{\bibinfo{journal}{arXiv e-prints}}
  \bibinfo{pages}{arXiv:2006.13528} (\bibinfo{year}{2020}).
\newblock \eprint{2006.13528}.

\end{thebibliography}

\normalsize
\vskip 0.2in
\noindent
{ACKNOWLEDGEMENTS.} We thank Manasvi Lingam for helpful comments on the manuscript. This work was supported in part by the Origins of Life Summer Undergraduate Research Prize Award and a grant from the Breakthrough Prize Foundation.

\vskip 1in

%\vskip 1in

\end{document}